\documentclass[12pt]{article}
\usepackage{fullpage,citesort,epsfig,graphics,amsbsy,amssymb}

\newcommand{\beq}{\begin{equation}}
\newcommand{\eeq}{\end{equation}}
\newcommand{\bea}{\begin{eqnarray}}
\newcommand{\eea}{\end{eqnarray}}

\newcommand\balpha{\mbox{\boldmath$\alpha$}}
\newcommand\bgamma{\mbox{\boldmath$\gamma$}}
\newcommand\bsigma{\mbox{\boldmath$\sigma$}}
\newcommand\bepsilon{\hbox{\twelvembf\char\number 15}}
\newcommand\Nlarge{N_c\rightarrow\infty}

\newcommand{\Tr}{{\rm Tr}}
\newcommand{\be}{\begin{equation}}
\newcommand{\ee}{\end{equation}}
\newcommand{\bq}{\begin{eqnarray}}
\newcommand{\eq}{\end{eqnarray}}
\newcommand{\ket}[1]{|#1\rangle}

\newcommand{\ie}{{\it i.e.\ }}
\def\math{\mathsurround=0pt }
\def\leftrightarrowfill{$\math \mathord\leftarrow \mkern-6mu
 \cleaders\hbox{$\mkern-2mu \mathord- \mkern-2mu$}\hfill
 \mkern-6mu \mathord\rightarrow$}
\def\overleftrightarrow#1{\vbox{\ialign{##\crcr
     \leftrightarrowfill\crcr\noalign{\kern-1pt\nointerlineskip}
     $\hfil\displaystyle{#1}\hfil$\crcr}}}

\newcommand{\VEV}[1]{\langle#1\rangle}

\let\l=\lambda

 \def\bd{\begin{document}} \def\ed{\end{document}}
\def\ds{\documentstyle} \let\fr=\frac \let\bl=\bigl \let\br=\bigr
\let\Br=\Bigr \let\Bl=\Bigl
\let\bm=\bibitem
\let\na=\nabla
\let\pa=\partial \let\ov=\overline
\def\ft#1#2{{\textstyle{{\scriptstyle #1}\over {\scriptstyle #2}}}}
\def\fft#1#2{{#1 \over #2}}
\def\vp{\varphi}
\def\sst#1{{\scriptscriptstyle #1}}
\def\oneone{\rlap 1\mkern4mu{\rm l}}
\def\td{\tilde}
\def\wtd{\widetilde}
\def\dalemb#1#2{{\vbox{\hrule height .#2pt
        \hbox{\vrule width.#2pt height#1pt \kern#1pt
                \vrule width.#2pt}
        \hrule height.#2pt}}}
\def\square{\mathord{\dalemb{6.8}{7}\hbox{\hskip1pt}}}
\def\wtd{\widetilde}
\def\R{\rlap{\rm I}\mkern3mu{\rm R}}
\def\im{{\rm i}}
\def\tilg{\tilde{g}}
\def\tilF{\tilde{F}}
\def\tilA{\tilde{A}}
\def\varf{\varphi}
\def\tilf{\tilde{\phi}}
\def\tilh{\tilde{h}}
\def\rme{{\rm e}}
\def\ep{\epsilon}
\def\0{{(0)}}
\def\9{{(9)}}
\def\8{{(8)}}
\def\7{{(7)}}
\def\6{{(6)}}
\def\5{{(5)}}
\def\4{{(4)}}
\def\3{{(3)}}
\def\2{{(2)}}
\def\1{{(1)}}
\newcommand{\trace}{{\rm Tr}}
\newcommand{\ub}{\overline{U}}
\newcommand{\vb}{\overline{V}}
\newcommand{\uh}{\widehat{U}}
\newcommand{\vh}{\widehat{V}}
\newcommand{\ubh}{\overline{\widehat{U}}}
\newcommand{\vbh}{\overline{\widehat{V}}}
\newcommand{\lb}{\bar{\l}}
\newcommand{\Fb}{\overline{F}}
\newcommand{\Fh}{\widehat{F}}
\newcommand{\Fbh}{\overline{\widehat{F}}}
\newcommand{\Ab}{\overline{A}}
\newcommand{\Ah}{\widehat{A}}
\newcommand{\Abh}{\overline{\widehat{A}}}
\newcommand{\Gb}{\overline{G}}
\newcommand{\Gh}{\widehat{G}}
\newcommand{\Gbh}{\overline{\widehat{G}}}
\newcommand{\Pb}{\overline{P}}
\newcommand{\Ph}{\widehat{P}}
\newcommand{\Pbh}{\overline{\widehat{P}}}
\newcommand{\Qb}{\overline{Q}}
\newcommand{\Qh}{\widehat{Q}}
\newcommand{\Qbh}{\overline{\widehat{Q}}}
\newcommand{\Bb}{\overline{B}}
\newcommand{\Bh}{\widehat{B}}
\newcommand{\Bbh}{\overline{\widehat{B}}}
\newcommand{\fhns}{\hat{F}^{\rm (NS)}}
\newcommand{\fhrr}{\hat{F}^{\rm (RR)}}
\newcommand{\ahns}{\hat{A}^{\rm (NS)}}
\newcommand{\ahrr}{\hat{A}^{\rm (RR)}}
\newcommand{\hhrr}{\hat{H}^{\rm (RR)}}
\newcommand{\hchi}{\hat{\chi}}
\newcommand{\hphi}{\hat{\phi}}
\newcommand{\htau}{\hat{\tau}}
\newcommand{\cG}{{\cal G}}
\newcommand{\cGb}{\overline{{\cal G}}}
\newcommand{\cH}{{\cal H}}
\newcommand{\cP}{{\cal P}}
\newcommand{\cPb}{\overline{{\cal P}}}
\newcommand{\cQ}{{\cal Q}}
\newcommand{\cQb}{\overline{{\cal Q}}}
\newcommand{\cM}{{\cal M}}
\newcommand{\cN}{{\cal N}}
\newcommand{\cO}{{\cal O}}
\newcommand{\cD}{{\cal D}}
\newcommand{\cL}{{\cal L}}

\newcommand{\vpp}{\mbox{$\langle{\scriptstyle++}\rangle$}}
\newcommand{\vmp}{\mbox{$\langle{\scriptstyle-+}\rangle$}}
\newcommand{\vppp}{\mbox{$\langle{\scriptstyle+++}\rangle$}}
\newcommand{\vmpp}{\mbox{$\langle{\scriptstyle-++}\rangle$}}
\newcommand{\vpmp}{\mbox{$\langle{\scriptstyle+-+}\rangle$}}
\def\balpha{\mbox{\boldmath$\alpha$}}
\def\bgamma{\mbox{\boldmath$\gamma$}}
\def\bsigma{\mbox{\boldmath$\sigma$}}
\def\bepsilon{\hbox{\twelvembf\char\number 15}}
\def\Nlarge{N_c\rightarrow\infty}
\def\Tr{{\rm Tr}}
\def\ie{{\it i.e.\ }}
\def\m@th{\mathsurround=0pt }
\def\leftrightarrowfill{$\m@th \mathord\leftarrow \mkern-6mu
 \cleaders\hbox{$\mkern-2mu \mathord- \mkern-2mu$}\hfill
 \mkern-6mu \mathord\rightarrow$}
\def\overleftrightarrow#1{\vbox{\ialign{##\crcr
     \leftrightarrowfill\crcr\noalign{\kern-1pt\nointerlineskip}
     $\hfil\displaystyle{#1}\hfil$\crcr}}}
\def\VEV#1{\langle#1\rangle}
\def\phdag{{\phantom{\dagger}}}
\begin{document}
\renewcommand{\thefootnote}{\fnsymbol{footnote}}
\begin{titlepage}
\phantom{.}

\vskip 3cm

\begin{center}
\begin{Large}
{\bf 6-Vertex Model on an Open String Worldsheet}
\end{Large}

\vskip 2.cm

{\large Charles B. Thorn\footnote{E-mail  address: thorn@phys.ufl.edu}}

\vskip 0.5cm

{\it Institute for Fundamental Theory,\\
Department of Physics, University of Florida,
Gainesville, FL 32611}

\vskip .5cm
\end{center}

\begin{abstract}
\noindent 
We propose boundary conditions on a two dimensional 6-vertex model, 
which is defined on the lightcone  
lattice for an open string worldsheet. We show that, in the continuum limit,
the degrees of freedom of this 6-vertex model 
describe a target space coordinate compactified on a circle of radius
$R$, which is related to the vertex weights. This conclusion 
had already been established for the
case of a 6-vertex model on the worldsheet lattice for the
propagator of a closed string. This exercise illustrates how the
Bethe {\it ansatz} works in the presence of boundaries, at least of
this particular type.
\end{abstract}
\end{titlepage}
\section{Introduction}
The lightcone worldsheet \cite{goddardgrt} lattice \cite{gilest} provides a
useful tool for analyzing the sum of planar diagrams
in field theory \cite{bardakcit,thornqcdsheet,thornsusysheet}
as well as in open string theory 
\cite{gilest,thornbits,papathanasioutclosed}. 
Once one commits to a lattice definition
of the worldsheet theory, it no longer is necessary to limit worldsheet
degrees of freedom to discretized versions of the continuum worldsheet
fields. For example, the worldsheet fermion fields in the Ramond-Neveu-Schwarz
model \cite{ramond,neveuschwarz}
can be represented on the lattice by Ising spin variables 
\cite{thornnsising}. One benefit of doing this is to eliminate the 
lattice fermion doubling problem.

In this spirit, one can use a 6-vertex model \cite{lieb,liebmw}
defined on the worldsheet lattice to realize a bosonic 
target space coordinate compactified on a circle \cite{gilesmt}.
In this context it is convenient to replace the standard
rectangular lattice discussed in \cite{gilesmt} 
by the diamond lattice arising in certain
fishnet models of the worldsheet \cite{nielsenfishnet,thornfishnet}. 
For the closed string worldsheet
this adaptation of the six vertex model to a diamond lattice was 
carried out in \cite{thorn6vnets},
where the continuum limit was carefully analyzed via the
Bethe {\it ansatz} \cite{bethe}, and its connection
to a compactified target space coordinate established. In this
short article we include boundaries in the 6-vertex model to establish
its equivalence in the continuum limit to a compactified
target space coordinate on the open string worldsheet.

It turns out that the mathematical analysis of the periodic case
discussed in \cite{thorn6vnets}, which closely follows
the work of Bethe \cite{bethe} and Yang and Yang \cite{yangyang}
on one dimensional Heisenberg spin chains and the work of \cite{lieb,liebmw}
on six vertex models, can be easily adapted
to include the case with boundaries of interest here.

We present our definition of the six-vertex model on an open 
string diamond lattice 
in the next section 2. In Section 3 we obtain the transfer matrix 
and construct its eigenstates using the Bethe {\it ansatz}.
In section 4 we analyze the eigenvalue spectrum of this transfer matrix
in the continuum limit We conclude with comments and discussion 
in Section 5.
In an appendix, we directly analyze our model for
the special value of the vertex
weight for which the transfer matrix is diagonalized 
by the states of a free fermion system. 
The availibility of explicit formulas in this case provides
an insightful confirmation of the results obtained in the main text.

\section{Six Vertex Model on a Worldsheet Lattice}
\begin{figure}[ht]
\begin{center}
\includegraphics[height=3.0in]{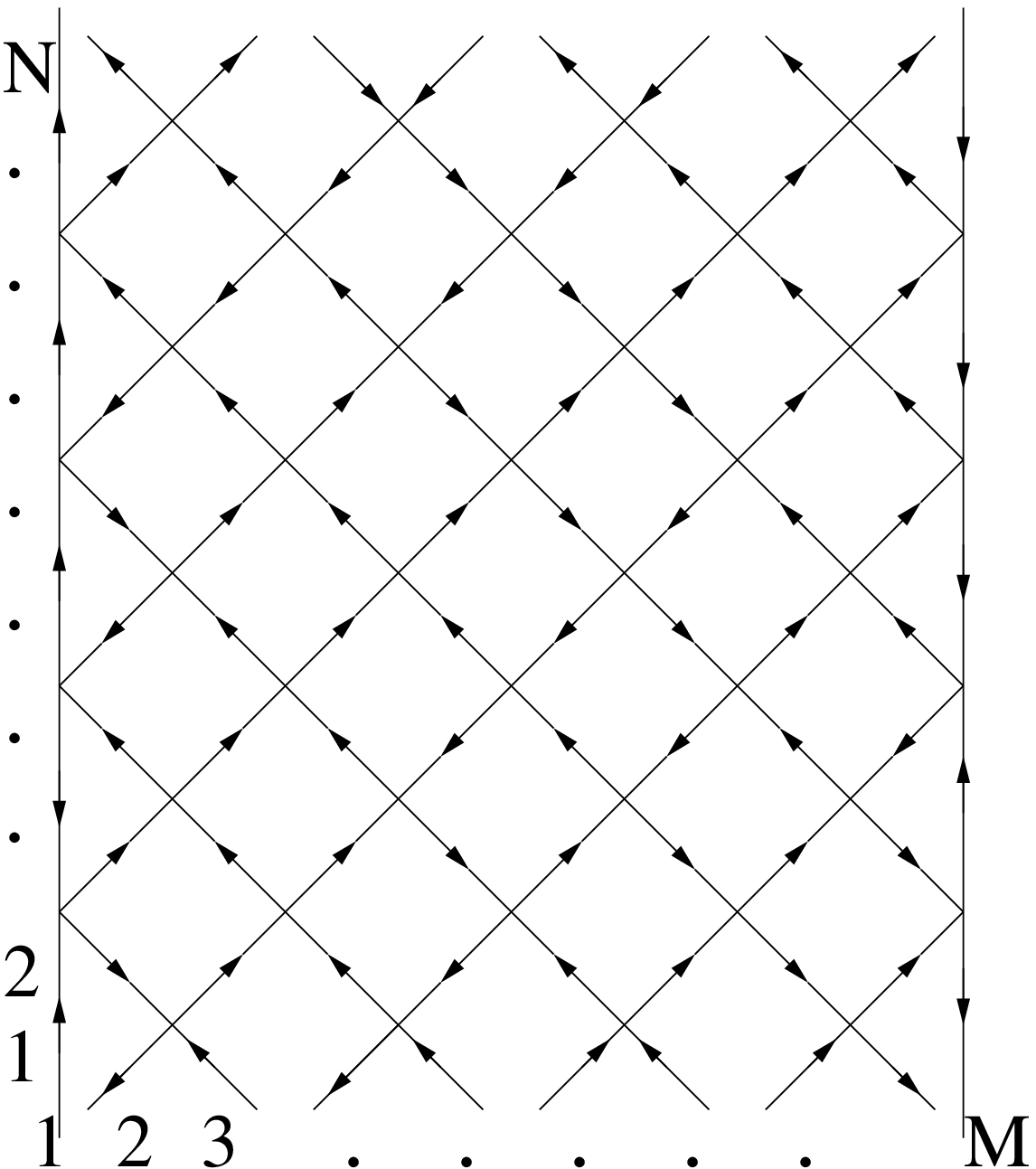}
\hskip18pt
\includegraphics[height=3.0in]{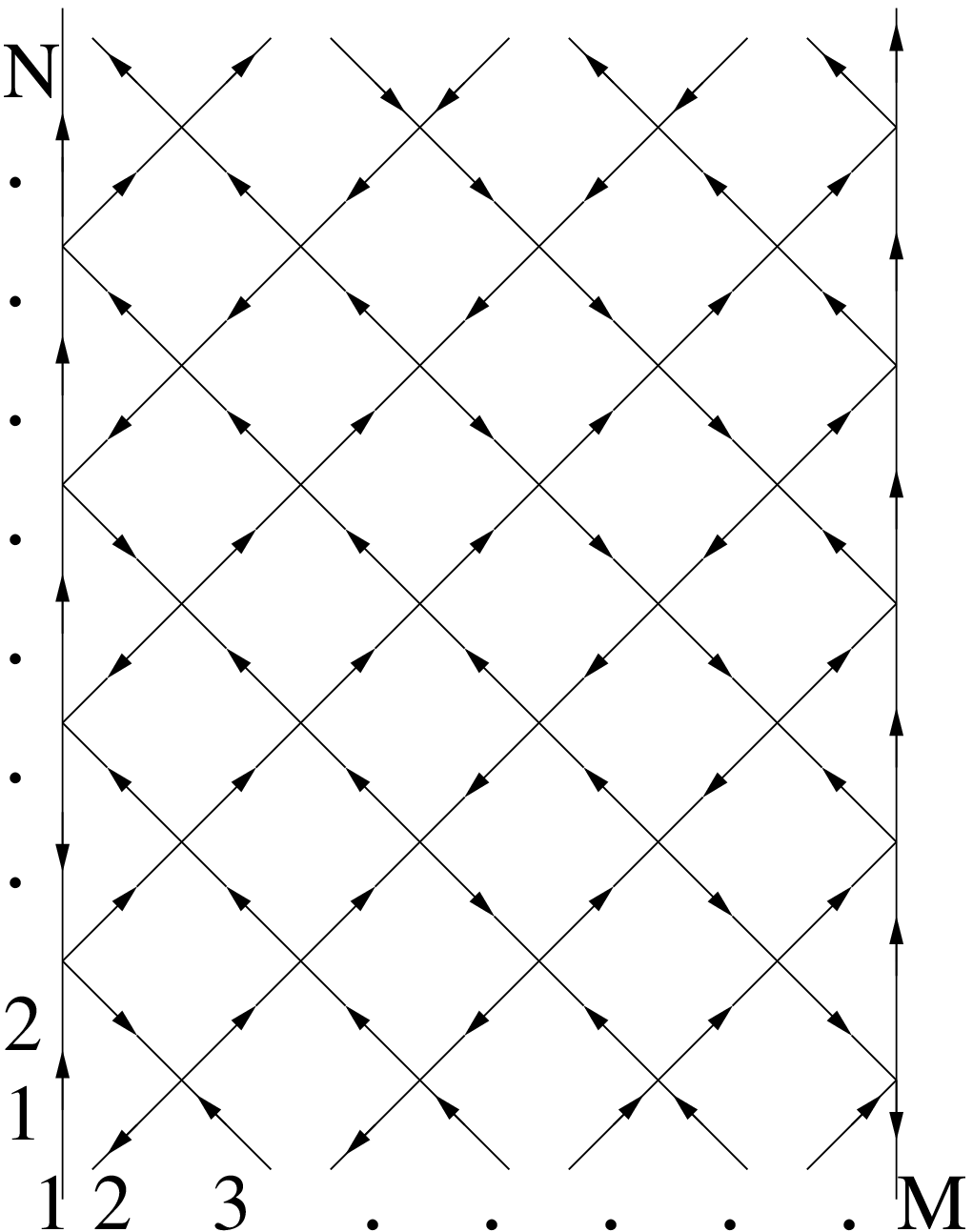}
\caption{Diamond worldsheet lattice propagating $M$ units of $P^+$ $N$ steps
in time. The left (right) figure shows even (odd) $M$.
Charge conserving
boundary conditions have been imposed.}
\label{6vevenodd}
\end{center}
\end{figure}

\begin{figure}[ht]
\begin{center}
\includegraphics[height=1.5in]{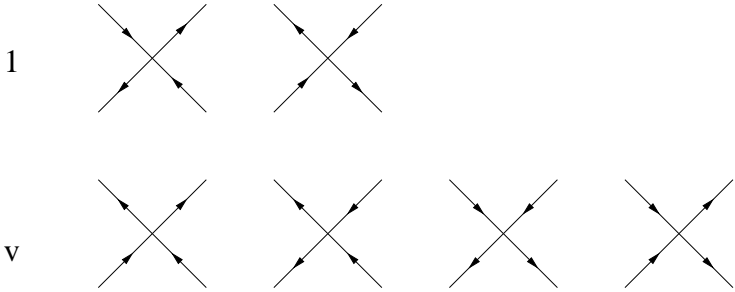}
\caption{The six charge conserving vertices.}
\end{center}
\label{6vertices}
\end{figure}

In this article we discuss the 6-vertex model on a diamond lattice,
illustrated with charge conserving
boundary conditions on the
vertical boundaries in Fig.\ref{6vevenodd}. In the worldsheet
interpretation we think of time as running vertically, and the
arrows at the bottom and top of the lattice specify possible
initial and final configurations of a worldsheet spin variable. 
Each link on this
lattice contains an arrow, which can be thought of as specifying
the direction of charge flow, each link carrying $\pm1$ unit
of charge.
There are precisely
six (planar) charge conserving vertices (see Fig.~\ref{6vertices}): 
Two with weight $1$ 
in which each adjacent
pair carries charge 0 into the vertex, and four with weight
$v$ in which two adjacent
lines carry charge 2 into the vertex. 
A typical fishnet diagram with these vertices is shown in
Fig.~\ref{6vevenodd}. The sum of all allowed arrow configurations is
thus seen to be equivalent to calculating the 
partition function for a 6-vertex model on a diamond
lattice.

To set up a six vertex model that is suitable for an open string
worldsheet, it is important that the charge $Q$ whose flow is
given by the vertex arrows is conserved at the boundaries. This
is necessary if $Q$ is to be identified as the zero mode momentum
of a compactified bosonic coordinate. A very natural choice is
shown on the left of Fig.~\ref{6vevenodd} for even $M$
and on the right of Fig.~\ref{6vevenodd} for odd $M$. 
Notice that in the even case the vertices
at the two boundaries are in step with each other, whereas in the 
odd case they are offset by one lattice step in time.
These figures show how to define the transfer matrix element 
between the arrow configurations at the top and bottom of each
figure. Just as in the case of periodic boundary conditions,
it is natural with the diamond lattice, to define the fundamental 
transfer matrix to cover two steps in time.

\section{The Transfer Matrix and its Eigenvalues}
The worldsheet lattice can be thought of as a discrete (imaginary) time
evolution of a state which is a tensor product of $M$ two state systems,
(``spins''), labeled by up and down arrows. Because of the
diamond lattice configuration, the basic discrete evolution is two
time steps, and we define each element of the
$2^M\times2^M$ transfer matrix ${\cal T}$ as the product of
vertex factors associated with the subgraph that connects
a given row of arrows with the row two time steps above it.
It is easy to see that the state with all arrows up or all arrows
down is an eigenstate of the transfer matrix with
eigenvalue $v^{M-1}$ for an open string worldsheet and $v^M$ for
a closed string worldsheet.
Because the transfer matrix conserves $Q$, we can work out its eigenstates
independently in each charge sector. The state with all arrows up
is the unique state with $Q=M$, and so it is automatically an eigenstate
with eigenvalue $v^{M-1}$, compared to $v^M$ for the analogous state with
periodic boundary conditions. The difference is explained by the
fact that with open string boundaries there is one less vertex in
two time steps.
\subsection{One overturned arrow ($Q=M-2$)}
With one overturned arrow we can label each state by the location
of that arrow $\ket{j}$, $j=1,\ldots,M$. If $j$ is sufficiently
far from both boundaries, in the bulk bulk of the worldsheet, 
the action of 
the transfer matrix is identical to that for periodic boundary
conditions \cite{thorn6vnets}, but with one
less power of $v$ on the right:
\begin{eqnarray}
{\cal T}\ket{j}=\cases{\ket{j+2}v^{M-1}+\ket{j+1}v^{M-2}+\ket{j-1}v^{M-2}
+\ket{j}v^{M-3}& for $j$ odd\cr
\ket{j-2}v^{M-1}+\ket{j+1}v^{M-2}+\ket{j-1}v^{M-2}
+\ket{j}v^{M-3}& for $j$ even.\cr}
\label{transferbulkb}
\end{eqnarray}
By direct inspection, we find that the action of ${\cal T}$ on the states 
with the overturned arrow close to the left boundary, with
$j=1,2$, is given by: 
\begin{eqnarray}
{\cal T}\ket{1}&=&\ket{1}v^{M-2}+\ket{2}v^{M-2}+\ket{3}v^{M-1}\\
{\cal T}\ket{2}&=&\ket{1}v^{M-1}+\ket{2}v^{M-3}+\ket{3}v^{M-2}
\label{transferleft}
\end{eqnarray}
When the overturned arrow is close to the right boundary,
we need to consider separately the
cases of $M$ even and odd:
\bea
{\cal T}\ket{M}&=&\cases{\ket{M}v^{M-2}+\ket{M-1}v^{M-2}+\ket{M-2}v^{M-1}
& $M$ even\cr
\ket{M}v^{M-2}+\ket{M-1}v^{M-1}& $M$ odd\cr}\\
{\cal T}\ket{M-1}&=&\cases{\ket{M}v^{M-1}+\ket{M-1}v^{M-3}+\ket{M-2}v^{M-2}
& $M$ even\cr
\ket{M}v^{M-2}+\ket{M-1}v^{M-3}+\ket{M-2}v^{M-2}+\ket{M-3}v^{M-1}
& $M$ odd\cr}
\eea
Because the transfer matrix acts locally, we can diagonalize its action
in the bulk by the same spin wave construction as in the periodic case
\begin{eqnarray}
\ket{k}_0&=&\sum_{j~{\rm odd}}\ket{j}e^{ikj}
+\xi(k)\sum_{j~{\rm even}}\ket{j}e^{ikj}\\
\xi(k)&\equiv&iv\sin k+\sqrt{1-v^2\sin^2 k}
\label{kwaveb}
\end{eqnarray}
with the eigenvalue $v^{M-1}t(k)$ with
\bea
t(k)&=&\left(\cos k+{1\over v}\sqrt{1-v^2\sin^2 k}\right)^2,
\eea
and for $-\pi<k<\pi$, these are all independent.

The states (\ref{kwaveb}) do not diagonalize the action of ${\cal T}$
near the boundaries. But because $t(-k)=t(k)$
We can construct the eigenstates in the presence of these boundaries
by taking a linear combination of the states 
$\ket{k}_0$ for $k>0$ and $-k$.
\bea
\ket{k}&=&\ket{k}_0+\eta(k)\ket{-k}_0\\
&=&\ket{1}(e^{ik}+\eta(k)e^{-ik})
+\ket{2}(\xi(k)e^{2ik}+\eta(k)\xi(-k)e^{-2ik})+\cdots
\eea
Focusing first on the left boundary, we apply ${\cal T}$ to the first 
few terms and collect the coefficient
of $\ket{1}$ to give the relation
\bea
v^{M-1}t(k)&=&v^{M-2}
+v^{M-1}{(\xi(k)e^{2ik}+\eta(k)\xi(-k)e^{-2ik})\over(e^{ik}+\eta(k)e^{-ik})}\\
\eta(k)&=&-{(vt(k)-1)e^{ik}-v\xi(k)e^{2ik}\over
(vt(k)-1)e^{-ik}-v\xi^*(k)e^{-2ik}}=-e^{2ik}
{vt-1-vz\over
vt-1-vz^*}
\eea
where we have used the definition $z=\xi e^{ik}$. Next we note that
\bea
z+v&=&v+(\cos k+i\sin k)(\sqrt{1-v^2\sin^2k}+iv\sin k)\nonumber\\
&=&(v\cos k+\sqrt{1-v^2\sin^2k})i\sin k +v\cos^2k+\cos k\nonumber\\
&=&(v\cos k+\sqrt{1-v^2\sin^2k})e^{ik}\\
e^{2ik}&=&{v+z\over v+z^*}\eea
and we rewrite
\bea
t(k)=\left(\cos k+{1\over v}(\xi-iv\sin k)\right)^2
=\left(e^{-ik}+{\xi\over v}\right)^2={1\over v^2}e^{-2ik}(v+z)^2
={(v+z)(v+z^*)\over v^2}\eea
From these relations we see that $\eta$ simplifies:
\bea
\eta(k)&=&-\left({v+z\over v+z^*}\right){(v+z)(v+z^*)-vz(v+z^*)
\over(v+z)(v+z^*)-vz^*(v+z)}=-{v+(1-v)z\over v+(1-v)z^*}
\eea
On the other hand we can also determine $\eta$ by applying ${\cal T}$
to the first few terms on the right of the row of arrows
\bea
&&\hskip-.75in\ket{k}_o=\ket{M}(e^{iMk}+\eta e^{-iMk})
+\ket{M-2}(e^{i(M-2)k}+\eta e^{-i(M-2)k})\nonumber\\
&&+\ket{M-1}
(\xi(k)e^{i(M-1)k}+\eta(k)\xi^*(k)e^{-i(M-1)k})+\cdots, \qquad M~{\rm odd}
\eea
and
\bea
&&\hskip-.75in\ket{k}_o=\ket{M}(\xi(k)e^{iMk}+\eta\xi^*(k) e^{-iMk})\nonumber\\
&&+\ket{M-1}
(e^{i(M-1)k}+\eta(k)e^{-i(M-1)k})+\cdots, \qquad M~{\rm even}
\eea
After applying ${\cal T}$ and collecting the coefficient of $\ket{M}$,
we obtain the relation for $M$ odd:
\bea
vt&=&1+v{e^{i(M-2)k}+\eta e^{-i(M-2)k}\over
e^{iMk}+\eta e^{-iMk}}+
{\xi(k)e^{i(M-1)k}+\eta(k)\xi^*(k)e^{-i(M-1)k}\over
e^{iMk}+\eta e^{-iMk}}\nonumber\\
\eta(k)&=&-{(vt-1)e^{iMk}-ve^{i(M-2)k}-\xi e^{i(M-1)k}\over
(vt-1)e^{-iMk}-ve^{-i(M-2)k}-\xi^* e^{-i(M-1)k}}\nonumber\\
&=&-e^{2iMk}{vt-1-e^{-2ik}(v+z)\over
vt-1-e^{2ik}(v+z^*)}=-e^{2iMk}{(v+z)(v+z^*)/v-1-(v+z^*)\over
(v+z)(v+z^*)/v-1-(v+z)}\nonumber\\
&=&-e^{2iMk}{z-1+1/v\over
z^*-1+1/v}={z\over z^*}\eta^*e^{2iMk}\eea
Here we assumed that $M$ was odd. The same result is obtained
for $M$ even, though the details differ:
\bea
vt&=&1+v{e^{i(M-1)k}+\eta(k)e^{-i(M-1)k}\over
\xi(k)e^{iMk}+\eta\xi^*(k) e^{-iMk}}\\
\eta(k)&=&-{(vt-1)\xi(k)e^{iMk}-ve^{i(M-1)k}\over
(vt-1)\xi^*(k)e^{-iMk}-ve^{-i(M-1)k}}=-{\xi\over\xi^*}
e^{2iMk}{vt-z^*(v+z)\over
vt-z(v+z^*)}\nonumber\\
&=&-{\xi\over\xi^*}{v+z\over v+z^*}
e^{2iMk}{(v+z^*)/v-z^*\over
(v+z)/v-z}={z\over z^*}\eta^*e^{2iMk}
\eea
which is identical to the result obtained with $M$ odd. 
Thus for all $M$ even and odd the quantization of $k$ is given by the
condition
\bea
e^{2iMk}&=&{z^*\over z}\eta^2(k),\qquad \eta(k)={v+(1-v)z\over v+(1-v)z^*}.
\eea
For periodic boundary conditions the quantization condition was the
much simpler $e^{iMk}=1$.

\subsection{$q$ overturned arrows ($Q=M-2q$)}
Eigenstates with several overturned arrows, in the presence
of boundaries, can again
be constructed by taking linear 
combinations of the Bethe {\it ansatz} in the bulk
which for $q=2$ is given by 
\bea
\ket{k_1,k_2}_0&=&\sum_{l<m}\ket{l,m}\left(\xi_l(k_1)\xi_m(k_2)e^{ilk_1+imk_2}
+A(k_1,k_2)\xi_l(k_2)\xi_m(k_1)e^{ilk_2+imk_1}\right)\\
A(k_1,k_2)&=&-{(1-1/v^2)z_2-z_1-z_1z_2/v-1/v\over
    (1-1/v^2)z_1-z_2-z_1z_2/v-1/v},
\eea
To economize notation we have affixed
a subscript to $\xi(k)\to \xi_l(k)$ such that $\xi_l(k)=1$ if $l$ is
odd and $\xi_l(k)=\xi(k)$ if $l$ is even.
Also we have defined $z_j\equiv \xi(k_j)e^{ik_j}$.
For $q$ overturned arrows, the Bethe {\it ansatz} is  
a sum over all permutations of the down arrows, and
$A(1,2)$ replaced by an $A_P$ for each permutation.
$A_P$ factors into a product of $A(k,l)$ for each pair interchange
needed to accomplish the permutation. When all down arrows are away from the
boundaries, the action of ${\cal T}$ diagonalizes on these bulk states
determining the eigenvalue of the transfer matrix to be
\begin{eqnarray}
T(k_1,\ldots,k_q)=v^{M-1}\prod_{j=1}^q t(k_j).
\end{eqnarray}
Since $T$ is invariant under the reversal of any of the $k_j\to-k_j$,
we can take linear combinations with each distinct term having 
one or more of the $k$'s reversed to diagonalize the action of
${\cal T}$ near the boundaries.

All of the essential features are already contained in the case
$q=2$, which we next analyze in detail, quoting the general
result at the end.  
Fixing $k_2$ for the moment we see by inspection that the combination
\bea
\ket{\psi_1}&=&\ket{k_1,k_2}+\eta(k_1)\ket{-k_1,k_2}
\eea
will properly realize the boundary conditions on the left for the
arrow associated with $k_1$, when it is to the left of that associated with
$k_2$. When the order of the down arrows is reversed, as in the second term,
the $k_1$ dependence is then
\bea
A(k_1,k_2)\xi_m(k_1)e^{imk_1}+\eta(k)A(-k_1,k_2)\xi_m(-k_1)e^{-imk_1}
&=&\nonumber\\
&&\hskip-3in A(k_1,k_2)\left[\xi_m(k_1)e^{imk_1}+\eta(k_1){A(-k_1,k_2)\over A(k_1,k_2)}
\xi_m(-k_1)e^{-imk_1}\right]
\eea
We see that the role of $\eta$ when the spin $k_1$ is on the left is
played by $\eta(k_1)A(-k_1,k_2)/A(k_1,k_2)=\eta(k_1)A(k_2,k_1)A(-k_2,k_1)$
when $k_1$ is on the right. It follows then that the boundary condition
on the right will be met by the $k_1$ arrow provided
\bea
e^{2iMk_1}&=&{z^*(k_1)\over z(k_1)}\eta^2(k_1)A(k_2,k_1)A(-k_2,k_1)
\eea
The symmetry of the $k_2$ dependence under $k_1\to-k_2$ is important
because it means that the construction 
\bea
\ket{\psi_2}&=&\ket{k_1.-k_2}+\eta(k_1)\ket{-k_1,-k_2}
\eea
leads to the same eigenvalue condition on $k_1$. To complete the construction
we need to form 
\bea
\ket{\psi}&=&\ket{\psi_1}+\eta(k_2){A(k_1,k_2)\over A_(k_1,-k_2)}\ket_{\psi_2}
\eea
which then satisfies the boundary conditions of the down arrow $k_2$
provided
\bea
e^{2iMk_2}&=&{z^*(k_2)\over z(k_2)}\eta^2(k_2)A(k_1,k_2)A(-k_1,k_2)
\eea
The generalization to any number $q$  of overturned spins is
now straightforward. The Bethe {\it ansatz} constructed along
parallel lines leads to the quantization conditions
\bea
e^{2iMk_r}&=&{z^*(k_r)\over z(k_r)}\eta^2(k_r)\prod_{s\neq r}
A(k_s,k_r)A(-k_s,k_r)
\label{openeigen}
\eea
We discuss the solution of these equations in the next section.
\section{Analysis of the Eigenvalue Equation}
The eigenvalue equation for the six-vertex model with boundaries
(\ref{openeigen}) can be cast in a form similar to the
eigenvalue equation with periodic boundary conditions analyzed in
\cite{thorn6vnets}. 
For comparison, recall that the eigenvalue equation in the periodic case
with $M^\prime$ arrows at each time slice, took the form
\begin{eqnarray}
e^{iM^\prime k_r}=\prod_{s\neq r}A(k_s,k_r).
\end{eqnarray}
To mimic the equation with boundaries we take $M^\prime=2M$, and take
$q^\prime=2q$ down arrows where half of them are associated with the $q$
$k_r>0$ in (\ref{openeigen}), and the other half are associated 
with the negatives $-k_r$ of these. Then the periodic equations take
the form
\bea
e^{2iM k_r}&=&A(-k_r,k_r)\prod_{s\neq r}A(k_s,k_r)A(-k_s,k_r),\qquad 
{\rm for}\quad k_r,k_s>0\\
e^{-2iM k_r}&=&A(k_r,-k_r)\prod_{s\neq r}A(k_s,-k_r)A(-k_s,-k_r)\nonumber\\
&=&A(-k_r,k_r)^*\prod_{s\neq r}A(-k_s,k_r)^*A(k_s,k_r)^*
\eea
The first equation is of the form (\ref{openeigen}) with
$\eta^2(k_r){z^*(k_r)/z(k_r)}$
replaced by $A(-k_r,k_r)$, and the second is the
complex conjugate of the first equation. Thus we can use the
results of \cite{thorn6vnets} to infer the continuum properties of the
system with boundaries. To do this we analyze the equation
\bea
e^{2iMk_r}&=&e^{i\Theta(k_r)}\prod_{s\neq r}A(k_s,k_r)
=-e^{i\Theta(k_r)+i\sum_{s\neq r}\theta(k_s,k_r)}\\
e^{i\Theta(k_r)}&=&-{z^*(k_r)\eta^2(k_r)\over z(k_r)A(-k_r,k_r)}
\eea
where $r,s=1,\ldots,2q$ and we constrain the solution to
satisfy $k_{2q+1-r}=-k_r$. The positive $k_r$'s will satisfy the
open eigenvalue equation. Taking the logarithm we can present the equation
to solve in the form
\bea
k_r&=&{\pi I_r\over M}+{\Theta(k_r)\over 2M}+{1\over2M}
\sum_{s\neq r}\theta(k_s,k_r)
\eea
where the $I_r$ are half odd integers satisfying the restriction
$I_{2q+1-r}=-I_r$. The second term on the right is the new feature
of the equations compared to those analyzed in \cite{yangyang,thorn6vnets}.
\subsection{The continuum limit $M\to\infty$}
For analyzing these equations  we map the $k_j$
onto new variables $\alpha_j$ for which $A(k_j,k_l)$ depends only on the
difference $\alpha_j-\alpha_l$. This is accomplished by the map
\cite{yangyang}
\begin{eqnarray}
z&=&\xi e^{ik}={e^{i\nu}-e^\alpha\over e^{i\nu+\alpha}-1}\nonumber\\
e^{i\nu}&=&{1\over 2v}+i\sqrt{1-{1\over4v^2}}.
\end{eqnarray}
Note that our parameter $\nu$ is related to a similar parameter $\mu=2\nu$
in \cite{yangyang}.
Here we restrict $\infty>v\geq1/2$, for which $e^{i\nu}$
is a pure phase. We note some special values of the mapping: 
$\alpha=0$ corresponds to $e^{ik}\xi=1$ which implies $k=0$, 
and $\alpha=\pm\infty$ map to $k=\pm(\pi-2\nu)$. 
(We are choosing $k$ to be in the 
range $-\pi<k<\pi$.) Thus the whole range $-\infty<\alpha<\infty$
corresponds to $-(\pi-2\nu)<k<\pi-2\nu$. Note that $v\to\infty$
shrinks the range of $k$ to 0, whereas $v\to1/2$ represents the
maximum range. It is straightforward to work out
the following quantities in terms of the new variables: 
\begin{eqnarray}
\tan k&=&{\sin2\nu\sinh\alpha\over\cos\nu-\cos2\nu\cosh\alpha}\nonumber\\
{dk\over d\alpha}&=&{\sin3\nu\over2[\cosh\alpha-\cos3\nu]}
+{\sin\nu\over2[\cosh\alpha-\cos\nu]}\nonumber\\
t(k)&=&\left(\cos k+{1\over v}\sqrt{1-v^2\sin^2 k}\right)^2
={\cosh\alpha-\cos3\nu\over\cosh\alpha-\cos\nu}\nonumber\\
A(k(\alpha),k(\beta))&=&-{1-e^{\beta-\alpha-4i\nu}\over e^{\beta-\alpha}-e^{-4i\nu}}
\equiv-e^{i\theta(\alpha,\beta)}\nonumber\\
\theta(\alpha,\beta)&=&2\arctan\left(\cot2\nu\tanh((\beta-\alpha)/2)\right)\\
\eta(k)&=&z{e^\alpha +e^{2i\nu}\over e^{\alpha+2i\nu}+1}\\
A(-k(\alpha),k(\alpha))&=&-{e^{4i\nu}-e^{2\alpha}\over
e^{4i\nu+2\alpha}-1}\\
e^{i\Theta(k(\alpha))}&=&{\sinh\alpha+i\sin2\nu\over
\sinh\alpha-i\sin2\nu},\qquad \Theta(\alpha)=2\arctan{\sin2\nu\over\sinh\alpha}
\end{eqnarray}
Using these equations, we can express the boundary conditions
in the alternative form
\begin{eqnarray}
k(\alpha_r)&=&{\pi I_r\over M}+{\Theta(\alpha_r)\over2M}
+{1\over 2M}\sum_{s\neq r}\theta(\alpha_s,
\alpha_r),
\end{eqnarray}
where the $I_l$ are half-odd integers since $q^\prime=2q$ is even. 
Different choices for these integers lead
to different solutions for the set of $k$'s, and hence
they provide us with a labeling of the eigenstates
of the transfer matrix. Yang and Yang \cite{yangyang}
have analyzed similar equations in their solution of the $x, y$
Heisenberg spin chain, and their techniques for solving them
in the limit $M\to\infty$ can be directly applied. 
For easy comparison,
we attempt as far as possible to adopt their notation. 
\subsection{The ground state with $Q^\prime=2Q>0$}
The ground state of the worldsheet system is the eigenstate
of the transfer matrix with maximal eigenvalue. For the problem with
periodic boundary conditions this state corresponds to the
choice of $I_r$'s symmetrically disposed about $0$, with no gaps
\cite{yangyang}.
For application to open boundary conditions, the symmetry about
$0$ is automatic due to the constraint on the $k_r$'s. Thus
the ground state corresponds to the choice
\bea
I_r&=&r-q-{1\over2},\qquad r=1,2,\ldots,2q
\eea
We remind the reader that our reference periodic system has
$M^\prime=2M$ arrows at each time step, $q^\prime=2q$ of which are
down. Thus the total charge of the reference system is $Q^\prime
=M^\prime-2q^\prime=2M-4q=2Q$ where $Q$ is the total charge of the
open system. In the reference periodic system the total
momentum $P^\prime=\sum_r k_r=(2\pi/M^\prime)\sum_r I_r$ is a
good quantum number which can be nonzero in general. But for 
application to the open system $P^\prime=0$ due to the symmetry of the
$k_r$ about $0$. Of course the actual open system has no conserved
momentum because of the presence of boundaries. 

We are interested in obtaining excitation energies of order $1/M$
above the ground state. If we try to calculate the total energy
of these states, we would have to not only calculate the $M^\prime\to\infty$
behavior of the energy, which is proportional to $M^\prime$, but also
corrections up to order $1/M^\prime$. However, excitation energies may be
obtained more simply by calculating energy differences $\Delta E=E(Q^\prime)
-E(0)$, as described in \cite{yangyang,thorn6vnets}. The trick is to
calculate $\Delta E$ in the thermodynamic limit $M\to\infty$ with $J=Q^\prime/M^\prime
=Q/M$ fixed and $P^\prime=0$. Of course for finite $Q^\prime$, 
we must examine the small
$J$ limit at the end of the calculation. We expect $\Delta E\propto
M^\prime J^2 =Q^{\prime2}/M^\prime$, which shows the desired $1/M^\prime$
dependence of the excitation energy.

In the thermodynamic limit the eigenvalue equation reduces to 
an integral equation for the density of eigenvalues $R(\alpha)$.
We define a kernel $K$ and density function
$R(\alpha)$ by
\begin{eqnarray}
K(\alpha,\beta)&\equiv&{1\over2\pi}{\partial\theta\over\partial\beta}
={1\over2\pi}{\sin4\nu\over\cosh(\alpha-\beta)-\cos4\nu}\nonumber\\
R(\alpha)&=&{2\pi\over M^\prime}{dj\over d\alpha},
\end{eqnarray}
and then the equation for the $k$'s as $M^\prime\to\infty$ 
becomes
\begin{eqnarray}
{dk\over d\alpha}&=&R(\alpha)+{1\over M^\prime}
{d\Theta\over d\alpha}+\int_{-\alpha_+}^{\alpha_+}{d\beta}
K(\alpha-\beta)R(\beta).
\end{eqnarray}
This equation has the same kernel $K$ as the one analyzed in 
\cite{thorn6vnets}, but the second term on the right is new to
the system with boundaries. However this new term vanishes in the
thermodynamic limit, so in the end, we can simply copy the results of
from this paper.

The value chosen for $\alpha_+$ determines the characteristics of the
eigenstate. For example, the eigenstate with maximum eigenvalue
$T^\prime$ for the transfer matrix corresponds to $\alpha_+=\infty$.
The values of $k$ at the limits of this range are $k=\pm(\pi-2\nu)$
and $t(k)=1$ for these values. As long as $0<\nu<\pi/2$,
$t(k)>1$ for all finite $\alpha$, so taking the whole range of
$\alpha$ corresponds to including in the expression for $T^\prime$
all values for $t$ greater than unity. For the continuum limit
we are only interested in very large $\alpha_+$ since then
the eigenvalues will be close (within $1/M^\prime$) of the maximum
eigenvalue.

From \cite{thorn6vnets} we quote $\Delta E^\prime$ of the reference
periodic model
\begin{eqnarray}
\Delta E^\prime&\sim&
{\pi-\mu\over4aM^\prime}{Q^{\prime2}}={\pi-\mu\over2aM}{Q^{2}}
\label{gaplessexc}
\end{eqnarray}
Because the reference periodic system has doubled the number of $k$'s,
this energy is twice the energy of the system with boundaries:
\bea
\Delta E&\sim&
{\pi-\mu\over4aM}{Q^{2}}={T_0\over 2P^+}\left[{\pi-\mu\over2}{Q^{2}}\right]
\eea
In brief, the charge dependence of the energy for the open system is
identical to that of the periodic system.

\subsection{Particle-hole excitations}
The particle-hole excitations in the reference periodic system also
correspond to excitations of the open system. In these excitations
the distribution of $I_r$'s is allowed to have gaps. Of course for
energies of order $1/M$, these gaps must be close to the ends
of the gapless distributions. For the periodic system the particle's
and holes near opposite ends of the gapless distribution can be
independently chosen. For the open system the constraint on the
$k_r$ requires that they always occur in equal and opposite
pairs of particles and holes. From \cite{thorn6vnets} we
quote the change in energy due to a particle-hole pair in the periodic system
\bea
\Delta E^\prime&=&{2\pi n\over M^\prime a}={\pi n\over M a}
\eea
where $n=|I_r-I_r^0|$ the integer $I_r^0$ has been replaced by the 
integer $I_r$. For the open system this excitation is matched
by one where $-I_r^0$ is replaced by $-I_r$. This doubles the
energy, but the energy of the open system is half the energy
of the reference periodic system so the energy change in the
open system is
\bea
\Delta E&=&{\pi n\over M a}=T_0{\pi n\over P^+}
\eea
For several particle-hole excitations $n_1,\ldots n_k$,
we simply replace $n$ by $N=\sum n_i$. Putting together all types of 
excitations we have the general expression for low-lying energy
eigenvalues
\bea
\Delta E&=&{T_0\over 2P^+}\left[{\pi-\mu\over2}Q^2+2\pi N\right]
\label{excitations}
\eea
\section{Discussion and Concluding Remarks}
Ref~\cite{thorn6vnets} established that in the periodic case the 
low lying spectrum of the six vertex model
matched that of a compactified coordinate on
the continuum closed string world-sheet,
described by the action
\begin{eqnarray}
S&=&{1\over2}\int d\tau\int_0^{P^+}d\sigma({\dot\phi}^2-T_0^2{\phi^\prime}^2)
\end{eqnarray}
with the equivalence relation
\begin{eqnarray}
\phi&\equiv& \phi+2\pi R.
\end{eqnarray}
This implies that the zero mode momentum conjugate to $\phi$ is quantized;
$p=k/R$ with $k$ an integer. The associated energy is $k^2/(2R^2P^+)$ 
There is an associated winding number $l$ for which
$\phi(p^+)-\phi(0)=2\pi lR$ which is associated with the energy
$4\pi^2l^2T_0^2R^2/(2P^+)$. Since $Q$ is even for the periodic case,
it is identified with $2k$. It then followed by comparison that 
$R^2=[2T_0(\pi-2\nu)]^{-1}$.

Now $\cos2\nu={\rm Re}~e^{2i\nu}=-1+1/2v^2$,
so the limit $R\to\infty$ implies $\nu\to\pi/2$ or $v\to\infty$.
The self dual radius $R_*^2=1/(2\pi T_0)$ corresponds to $\nu=0$
or $v=1/2$. Thus the range of couplings considered here
$1/2\leq v<\infty$ (for which the 6-vertex model is
critical) produces circle radii $R_*\leq R <\infty$.
Interestingly, small radii, $R<R_*$ are not accessible
in the vertex model. For $v<1/2$ the model is not critical
and the continuum limit accordingly sends all excitations
to infinite energy, i.e. there is no interesting continuum
limit.

In this article we have obtained the low lying spectrum for the
open string worldsheet lattice. There is of course no winding number,
but the $Q$ dependence of the energy is exactly as in the closed
string case, with the exception that $M$ can be odd,
in which case $Q$ is odd.
In the compactified coordinate interpretation, this implies
that, under the shift $\phi\to
\phi+2\pi R$, the wave function of the open string is periodic when
$M$ is even and antiperiodic when $M$ is odd.

We have therefore confirmed the expectation that the six vertex
model on the diamond lattice provides a satisfactory discretization
of a compactified target space coordinate for both
open and closed strings. This discretization
may be particularly effective in monte carlo simulations of
the sum of all planar diagrams of open string theory as advocated in
\cite{papathanasioutclosed}.

\vskip18pt
\underline{Acknowledgments:} 
This work is supported in part by U.S. Department of Energy
under grant DE-FG02-97ER-41029.
\appendix
\section{The free fermion case $v=1/\sqrt{2}$}
As a useful check on our conclusions, we study the case $v=1/\sqrt{2}$
($\nu=\pi/4$) for which $A=-1$. Then the quantization conditions on the $k_r$
decouple and reduce to
\bea
e^{2iMk_r}&=&{z^*(k_r)\over z(k_r)}\left({1+z(\sqrt{2}-1)
\over1-z^*(\sqrt{2}-1)}\right)^2,\qquad r=1,\ldots,q\\
z(k)&=&{e^{ik}\over\sqrt{2}}(i\sin k+\sqrt{1+\cos^2k})
\eea
Here there is no need to use a reference periodic system, and no need
to double the $k_r$'s. An eigenstate involves any number of overturned
spins which can be independently assigned a momentum solving
this equation. Just as with a free Fermi gas, the ground state
is obtained by populating all the $k_r$ with $t(k_r)>1$. Dropping
the index, we have, for $v=1/\sqrt{2}$,
\bea
t(k)&=&\left(\cos k+\sqrt{1+\cos^2k}\right)^2
\eea
and we see that $t(k)=1$ for $k=\pi/2$, and $t(k)>1$ for $k<\pi/2$.
The low lying excitations all arise from altering the population
of overturned spins with $k\approx\pi/2$, leaving the overturned
spins with $k-\pi/2$ of order unity in their ground state 
configuration. To study the spectrum of these low lying excitations,
put $k=\delta+\pi/2$. Then
\bea
t(k)&\to&\left(-\sin\delta+\sqrt{1+\sin^2\delta}\right)^2
\sim 1-2\delta+{\mathcal O}(\delta^2)\\
\Delta E&=&-{\ln t\over2a}\sim {\delta\over a}+{\mathcal O}(\delta^2)
\eea
Next we examine the quantization condition for $k\approx\pi/2$.
We find $z\to e^{3i\pi/4+i\delta}+{\mathcal O}(\delta^2)$ 
and the right side of the quantization condition becomes
\bea
{z^*(k)\over z(k)}\left({1+z(\sqrt{2}-1)
\over1-z^*(\sqrt{2}-1)}\right)^2&\to&-1+{\mathcal O}(\delta)
\eea
Then the quantization equation reads
\bea
e^{2iM(\delta+\pi/2)}&=&(-)^Me^{2iM\delta}=-1+{\mathcal O}(\delta)\\
\delta&=&{(2I+1)\pi\over2M}+{\mathcal O}(\delta/M),\qquad M\quad{\rm even}\\
\delta&=&{I\pi\over M}+{\mathcal O}(\delta/M),\qquad M\quad{\rm even}
\eea
where $I$ is any integer.
Since $\delta$ starts out at order $1/M$, it is safe to drop
the correction terms ${\mathcal O}(\delta/M)$ to this solution.
These results show immediately that particle-hole excitations,
which leave the number of overturned arrows constant, change the
energy by an integer multiple of $\pi/(Ma)$.

To compare energies in different charge sectors, we change the number
of overturned arrow. Start with the lowest energy state with
charge zero. This state requires that $M$ is even and there are
$q=M/2$ overturned arrows, all populating all the levels
with negative energy. We can increase the charge by $2n$ units
by flipping the $n\ll M$ arrows at the top of the sea. These previously
down arrows were contributing a negative energy, so flipping them
increases the energy by the amount
\bea
\Delta E(2n)&=&{\pi\over2Ma}\left[1+3+\cdots+(2n-1)\right]
={\pi\over2Ma}\left[n(n+1)-n\right]={\pi n^2\over2Ma}={\pi Q^2\over8Ma}
\eea
which agrees with our result (\ref{excitations}) for $\mu=2\nu=\pi/2$.
Of course, when $M$ is even only sectors with even charge can appear.

To reach odd values of the charge we need to take $M$ odd. In this
case there is no state of zero charge: the lowest energy states
has $Q=\pm1$. These two degenerate states are connected by
the overturned arrow with $k=0$, which is allowed when $M$ is
odd. So start with the $Q=1$ state. Then flipping $n\ll M$
arrows at the top of the sea reaches the state with
$Q=1+2n$. This increases the energy by
\bea
{\pi\over Ma}\left[1+2+\cdots+n\right]&=&{\pi n(n+1)\over2Ma}
={\pi(Q-1)(Q-1+2)\over8Ma}={\pi Q^2\over8Ma}-{\pi\over8Ma}
\eea
This is consistent with the result (\ref{excitations})
but leaves open the possibility that there is a $Q$ independent
shift in the energies between the cases with  even and odd $M$.
However, we know the $1/M$ contribution to the large $M$ behavior
of the ground state energy in any sector of a free fermion
system is determined by the well-known Casimir zero-point
energy calculation. When $M$ is even the low energy frequencies
are $(n+1/2)\pi/(aM)$ which is known to give a contribution
of $-d\pi/(48aM)$ where $d$ is the number of Fermi fields:
$d=2$ for the present case of free charged fermions. 
When $M$ is odd, the Casimir
zero-point energy is $+d\pi/(24aM)$. Thus the energy difference
between energies in the even and odd $M$ sectors has the $1/M$ dependence
\bea
{d\pi\over24aM}-{-d\pi\over48aM}&=&{d\pi\over16aM}\to{\pi\over8aM}
\eea
for $d=2$. The boundary terms must also match by locality. 
The bulk terms $\alpha M$ will have the same $\alpha$ for even
and odd $M$, but of course
$M$ itself will be different in even and odd sectors. So, 
as a consequence of these
general arguments we can conclude that the low-lying  energies 
are
\bea
E&=&\alpha M+\beta-{\pi\over24aM}+{\pi Q^2\over 8aM}+{\pi N\over aM}
\eea
where $N=\sum_l n_l$ is the total mode number of the particle hole excitations,
where the $n_l$ are nonnegative integers. We now understand that
$Q$ is even when $M$ is even and $Q$ is odd when $M$ is odd.


\begin{thebibliography}{10}

\bibitem{goddardgrt}
P. Goddard, J. Goldstone, C. Rebbi, and C. B. Thorn, {\sl Nucl. Phys.} {\bf
  B56} (1973) 109.
\bibitem{gilest}
R. Giles and C. B. Thorn, {\sl Phys. Rev.} {\bf D16} (1977) 366.

\bibitem{bardakcit}
  K.~Bardakci and C.~B.~Thorn,
  Nucl.\ Phys.\ B {\bf 626} (2002) 287
  [hep-th/0110301].

\bibitem{thornqcdsheet}
  C.~B.~Thorn,
  Nucl.\ Phys.\ B {\bf 637} (2002) 272
   [Erratum-ibid.\ B {\bf 648} (2003) 457]
  [hep-th/0203167].

\bibitem{thornsusysheet}
  S.~Gudmundsson, C.~B.~Thorn and T.~A.~Tran,
  Nucl.\ Phys.\ B {\bf 649} (2003) 3
  [hep-th/0209102].

\bibitem{thornbits}
  C.~B.~Thorn,
  In *Moscow 1991, Proceedings, Sakharov memorial lectures in physics, vol. 1* 447-453 [hep-th/9405069];
  Phys.\ Rev.\ D {\bf 80} (2009) 086010
  [arXiv:0906.3742 [hep-th]].

\bibitem{papathanasioutclosed}
  G.~Papathanasiou and C.~B.~Thorn,
  Phys.\ Rev.\ D {\bf 86} (2012) 066002
  [arXiv:1206.5554 [hep-th]];
  Phys.\ Rev.\ D {\bf 87} (2012) 066005
  [arXiv:1212.2900 [hep-th]];
  arXiv:1305.5850 [hep-th].
\bibitem{ramond}
  P.~Ramond,
  Phys.\ Rev.\  D {\bf 3} (1971) 2415;
A.~Neveu and J.~H.~Schwarz,  
 Phys.\ Rev.\  D {\bf 4} (1971) 1109;
  C.~B.~Thorn,
  Phys.\ Rev.\  D {\bf 4} (1971) 1112.
\bibitem{neveuschwarz}  A.~Neveu and J.~H.~Schwarz,
Nucl.\ Phys.\  B {\bf 31} (1971) 86;
  A.~Neveu, J.~H.~Schwarz and C.~B.~Thorn,
  Phys.\ Lett.\  B {\bf 35} (1971) 529.

\bibitem{thornnsising}
  C.~B.~Thorn,
  Phys.\ Rev.\ D {\bf 80} (2009) 086010
  [arXiv:0906.3742 [hep-th]].

\bibitem{lieb}
E. H. Lieb, {\sl Phys. Rev. Lett.} {\bf 18} (1967) 692; {\sl Phys. Rev.} {\bf
  162} (1967) 162.

\bibitem{liebmw}
E. H. Lieb, {\sl Phys. Rev. Lett.} {\bf 18} (1967) 1046; B. McCoy and T. T. Wu,
  {\sl Nuovo Cimento} {\bf 56B} (1968) 311; B. Sutherland, {\sl J. Math. Phys.}
  {\bf 11} (1970) 3183.

\bibitem{gilesmt}
R. Giles, L. McLerran, and C. B. Thorn, {\sl Phys. Rev.} {\bf D17} (1978) 2058.
\bibitem{nielsenfishnet}
H. B. Nielsen and P. Olesen, {\sl Phys. Lett.} {\bf 32B} (1970) 203; B. Sakita
  and M. A. Virasoro, {\sl Phys. Rev. Lett.} {\bf 24} (1970) 1146.

\bibitem{thornfishnet}
C. B. Thorn, {\sl Phys. Lett.} {\bf 70B} (1977) 85; {\sl Phys. Rev.} {\bf D17}
  (1978) 1073.

\bibitem{thorn6vnets}
  C.~B.~Thorn,
  Phys.\ Rev.\ D {\bf 63} (2001) 105009
  [hep-th/0012006].

\bibitem{bethe}
H. A. Bethe, {\sl Z. Phys.} {\bf 61} (1930) 206.

\bibitem{yangyang}
C. N. Yang and C. P. Yang, {\sl Phys. Rev.} {\bf 150} (1966) 321; {\sl Phys.
  Rev.} {\bf 150} (1966) 327.

\end{thebibliography}
\end{document}